\documentclass{IEEEtran}

\usepackage[
backend=biber,
style=ieee,
sorting=none 
]{biblatex}

\usepackage{amsmath,amssymb,amsfonts}
\usepackage{graphicx}
\usepackage{xcolor} 

\usepackage{textcomp,nicefrac}
\def\BibTeX{{\rm B\kern-.05em{\sc i\kern-.025em b}\kern-.08em
T\kern-.1667em\lower.7ex\hbox{E}\kern-.125emX}}
\begin{document}
\title{Experience and performance of persistent memory for the DUNE data acquisition system}
\author{Adam Abed Abud, Student Member, IEEE, Giovanna Lehmann Miotto, Roland Sipos
\thanks{This  work  has  been  submitted  to  the  IEEE  for  possible  publication.Copyright may be transferred without notice, after which this version may nolonger be accessible.} 
\thanks{Adam Abed Abud (corresponding author) is with both University of Liverpool, The Oliver Lodge, Oxford St, Liverpool L69 7ZE, United Kingdom and European Laboratory for Particle Physics (CERN), Geneva 23, CH-1211, Switzerland (e-mail: adam.abed.abud@cern.ch).}
\thanks{Giovanna Lehmann Miotto and Roland Sipos are with European Laboratory for Particle Physics (CERN), Geneva 23, CH-1211, Switzerland.}

}

\maketitle

\begin{abstract}

Emerging high-performance storage technologies are opening up the possibility of designing new distributed data acquisition system architectures, in which the live acquisition of data and their processing are decoupled through a storage element. An example of these technologies is 3DXPoint, which promises to fill the gap between memory and traditional storage and offers unprecedented high throughput for data persistency. 

In this paper, we characterize the performance of persistent memory devices, which use the 3DXPoint technology, in the context of the data acquisition system for one large Particle Physics experiment, DUNE. This experiment must be capable of storing, upon a specific signal, incoming data for up to 100 seconds, with a throughput of 1.5~TB/s, for an aggregate size of 150 TB. The modular nature of the apparatus allows splitting the problem into 150 identical units operating in parallel, each at 10~GB/s. The target is to be able to dedicate a single CPU to each of those units for data acquisition and storage.

\end{abstract}


\section{Introduction}
\label{sec:introduction}

Over the last years, large-scale computing systems are moving towards a direction where compute and storage capabilities are decoupled. In fact, high performance computing environments are experiencing a huge increase in data volumes which make the data consumption more difficult for the computing nodes. By decoupling compute and storage it is, therefore, possible to scale the system by increasing the storage hardware when more data is produced. 


As a consequence, emerging high performance storage technologies are being used in the design of new distributed data acquisition system architectures where data production and data processing are decoupled by a large storage buffer. An example of these technologies is 3DXPoint which promises to fill the gap between memory and traditional storage. Devices which are based on this recent technology provide a memory-like bandwidth and a large data capacity.

One possible application of 3DXPoint devices is in the context of the DUNE detector \cite{DUNE_detector}. This is a long baseline neutrino experiment due to take data in 2026. The data acquisition (DAQ) of the DUNE experiment will consist of a large-scale distributed system designed to handle a total of 1.5~TB/s of incoming data from the readout system. The DUNE baseline design takes advantage of 150 custom PCIe cards (FELIX readout from the ATLAS experiment at CERN \cite{FELIX_ATLAS}) that push data into the host memory at a data rate of 1~GB/s for each of the 10 optical links. The current system foresees two of these devices for each dual socket server. A high throughput storage application is needed to keep up with the incoming data once a trigger signal is detected. From an implementation point of view this translates into continuously transferring the data from the PCIe readout devices to the host's DRAM via \textit{direct memory access} (DMA) and then, at the start of the trigger signal, storing the the data from memory to persistent storage media. This has to be done continuously for approximately 100~s at a data rate of 10~GB/s for each CPU socket. As a consequence, the total storage needed will be at least 1~TB for each CPU.

Given the importance of the data being recorded, DRAM technology is not a viable solution for the DUNE storage buffer as it cannot provide storage persistence. In addition, the total buffer size needed for each server would make the system too costly with only DRAM modules. One possible way to achieve the target goal for the system is to use fast storage media such as the Intel\textregistered\, Optane\texttrademark\, Data Center Persistent Memory Modules (DCPMMs) which leverage the 3DXPoint technology. 

In this paper, we will assess the raw storage performance of DCPMMs. A careful evaluation was done focusing on both the throughput and the request rate. This was executed with both synthetic benchmarks and with a custom-made high level application. In a second step, the suitability of DCPMMs for the DUNE storage buffer was explored. Initially, this was done with a prototype setup available at CERN \cite{protoDUNE_DAQ} and subsequently with an application resembling as much as possible the required workload for a possible application within the DUNE experiment.

\section{Intel\textregistered\, Optane\texttrademark}

\subsection{3D XPoint\texttrademark\, memory technology}

Intel\textregistered\, Optane\texttrademark\, devices are based on the 3D XPoint\texttrademark\, memory technology. This is a new type of non-volatile technology that offers 10 times higher bandwidth and approximately 1000 times more endurance compared to traditional NAND based storage media. 3D XPoint\texttrademark\, should not be confused with 3D NAND. In the latter, several storage cells (transistors) are stacked on top of each other to form a single structure for better space utilization whereas in 3D XPoint\texttrademark\, data storage cells are treated independently. Therefore, contrary to NAND storage media, 3D XPoint\texttrademark\, devices are byte addressable. Manufacturers have produced both solid state drives and memory modules that use 3D XPoint\texttrademark\, technology. In this paper, we will focus on the performance of the Intel\textregistered\, Optane\texttrademark\, Data Center Persistent Memory Modules.

Non-volatile memory (NVM) is an emerging and innovative technology that sits on the memory bus, like commonly available DRAM modules. DCPMMs are non-volatile memory devices that use the 3D XPoint\texttrademark\, technology. They offer memory-like performance at a lower cost per gigabyte compared to DRAM. Therefore, DCPMMs are good candidates to fill the performance gap between memory and storage devices. In addition, contrary to storage media where data access is usually done with a 4~KiB block size, DCPMMs fetch the data with 4 cache lines of 64 bytes. This results in a 256 bytes load/store instruction that provides a lower latency, similar to memory devices.

\subsection{Operation Modes}

From an operational point of view DCPMMs can be configured into three modes: 

\begin{itemize}
    \item \textit{Memory mode}: in this mode the DCPMMs act as a large memory pool alongside the DDR4 memory modules. Therefore, DCPMMs are seen by the operating system as a large volatile memory pool.
    \item \textit{App Direct mode}: in this mode the DCPMMs provide in-memory persistence by acting as storage devices rather than memory devices. The memory controller maps the DCPMMs to the physical memory address space of the machine so that the software layer can directly access the devices. 
    \item \textit{Mixed mode}: in this mode it is possible to use a percentage of the DCPMMs capacity in both Memory and App Direct modes. 
    
\end{itemize}

\begin{figure}[t]
\centerline{\includegraphics[width=3.5in]{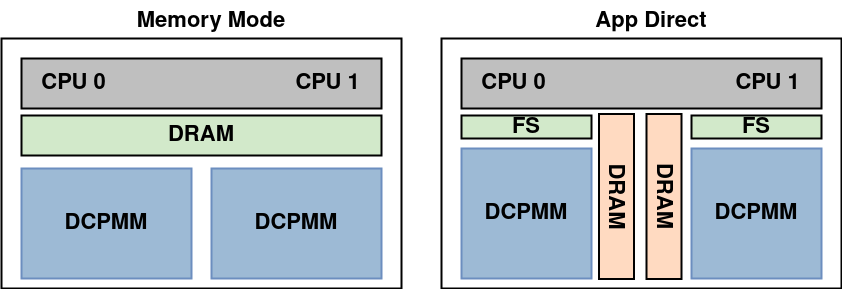}}
\caption{Logical view of both the the Memory and App Direct mode of operation for DCPMM devices.}
\label{fig:app_direct_memory}
\end{figure}

Fig. \ref{fig:app_direct_memory} shows a logical view of the two main modes of operational of DCPMMs.

In addition, in the App Direct mode the DCPMMs can be configured into two ways: 

\begin{itemize}
    \item \textit{Interleaved region}: all the DCPMMs relative to a CPU socket are seen as a single block device as if the modules are used in a RAID-0 configuration. 
    \item \textit{Non-interleaved region}: each DCPMM is seen a single block device. Therefore, each module is accessed independently.
    
\end{itemize}

Finally, depending on the configuration, it is possible to mount the DCPMMs with a  \textit{direct access} file system (DAX). This provides a byte-addressable access to the storage without the need to perform an extra copy on the page cache and, thus, it yields higher read and write bandwidths.

\section{Evaluation}
\subsection{System description}
Table \ref{testbed_setup} summarizes the specification of the machine node used for the evaluation. The test machine is a dual CPU socket system with 56 physical cores on each processor and one memory controller per socket. Each memory controller has 6 memory channels composed by both a DDR4 DRAM device (16~GB) and a DCPMM (512~GB). The total DRAM size of the machine is 192~GB whereas the total DCPMM size is 6~TB. Finally, the node is installed with a Centos 7 operating system and kernel version 4.15. 

\subsection{Testing strategy}

The raw performance of the persistent memory devices was obtained by executing a synthetic benchmark evaluation with the \textit{flexible-IO} (FIO) tool \cite{FIO}. This was executed with the DCPMMs used as a storage device in App Direct mode and mounted with a DAX-enabled ext4 file system. After the raw benchmarks, the system was also tested with a high level C++ application that used the DCPMMs as the target storage media. The setup under investigation has been tested as a function of both the block size and the number of threads. The resulting performance was obtained in terms of request rate and in terms of total throughput. 
It is also worth mentioning that the results obtained showed good reproducibility across several runs. However, the uncertainty on the single measurements was not computed as it was not part of the goals of the research. Note that the benchmarks executed on the system refer to the 6 DCPMMs connected to the same CPU socket, unless otherwise stated.


\begin{table}[]
\caption{Overview of the test machine used for evaluation.}
\label{testbed_setup}
{\renewcommand{\arraystretch}{1.5}
\normalsize %
\begin{tabular}{cl}
\hline
\textbf{CPU}   & \begin{tabular}[c]{@{}l@{}}Intel\textregistered\, Xeon\textregistered\, Platinum 8280L  \\ 2.70 GHz (Cascade Lake), dual socket\\ L1d cache 32K \\ L2 cache 1024K\\ L3 cache 3942K\end{tabular} \\ \hline
\textbf{DRAM}  & DDR4 DRAM 16 GB, 2666 MT/s, 12 slots                                                                                                             \\ \hline
\textbf{DCPMM} & DDR-T 512 GB, 2666 MT/s, 12 slots                                                                                                                \\ \hline
\textbf{OS}    & Centos 7, Linux Kernel 4.15.0                                       
  \\ \hline
\textbf{SW} & ipmctl v.01.00, PMDK v.1.9, fio v.3.2                   
\\ \hline
\end{tabular}
}
\end{table}

\section{Benchmnarks of the DCPMMs and discussion}
\label{benchmarks_DCPMMs}

\begin{figure}[t]
\centerline{\includegraphics[width=3.5in]{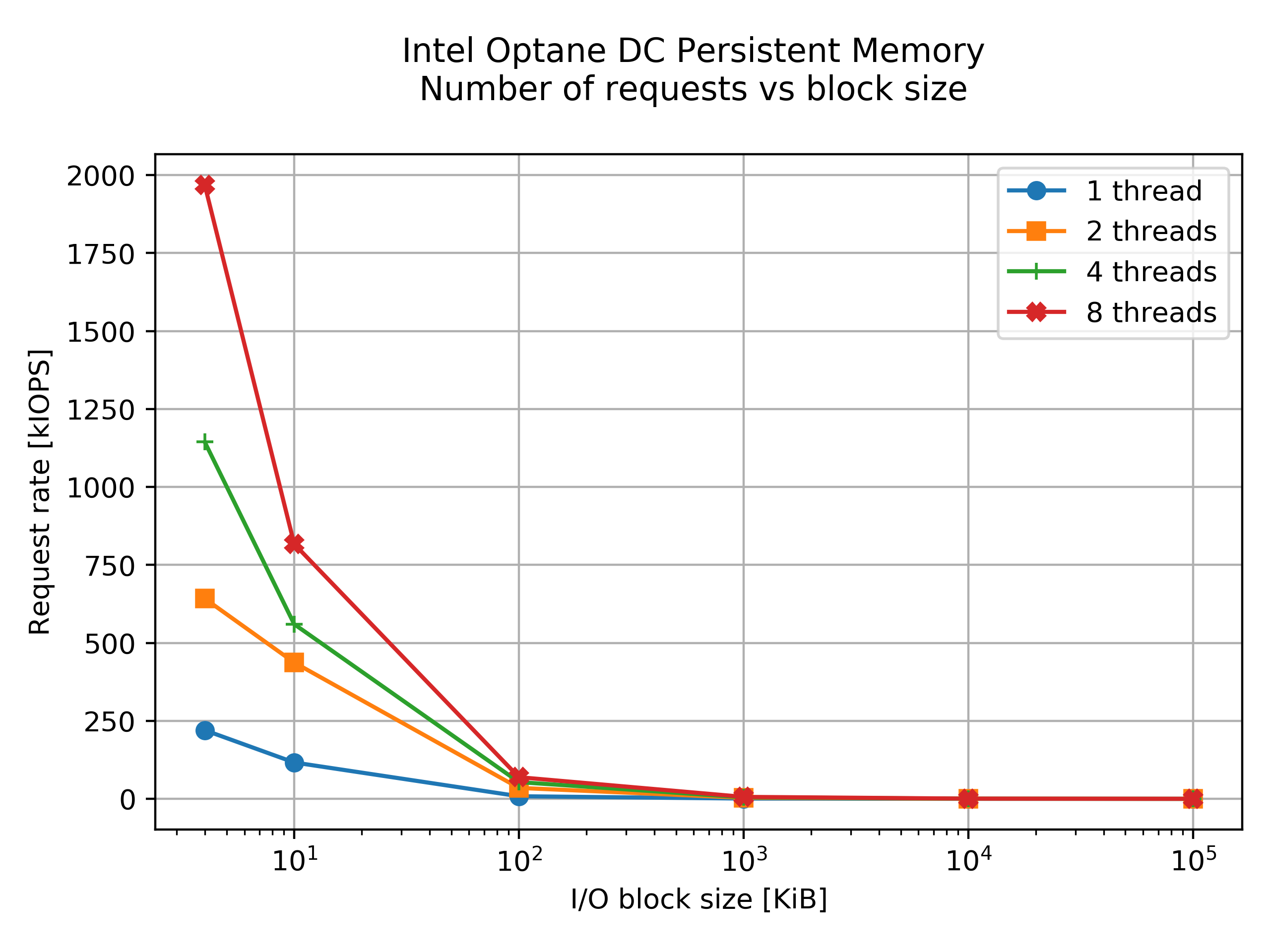}}
\caption{Request rate as a function of the I/O block size for different writing threads. Increasing the block size results in a smaller request rate due to the increased latency time to fetch and store the requested block.}
\label{fig:request_blocksize}
\end{figure}

Fig. \ref{fig:request_blocksize} represents the request rate as a function of the I/O block size for different writing threads. This was obtained by measuring the number I/O operations per second for a given workload in terms of block size and number of threads. Increasing the block size results in a smaller request rate due to the increased latency time to fetch and store the requested block. The maximum rate sustained by a single writing thread is approximately 200k operations per second. This showcases the small time needed to request data with DCPMMs.



\begin{figure}[t]
\centerline{\includegraphics[width=3.5in]{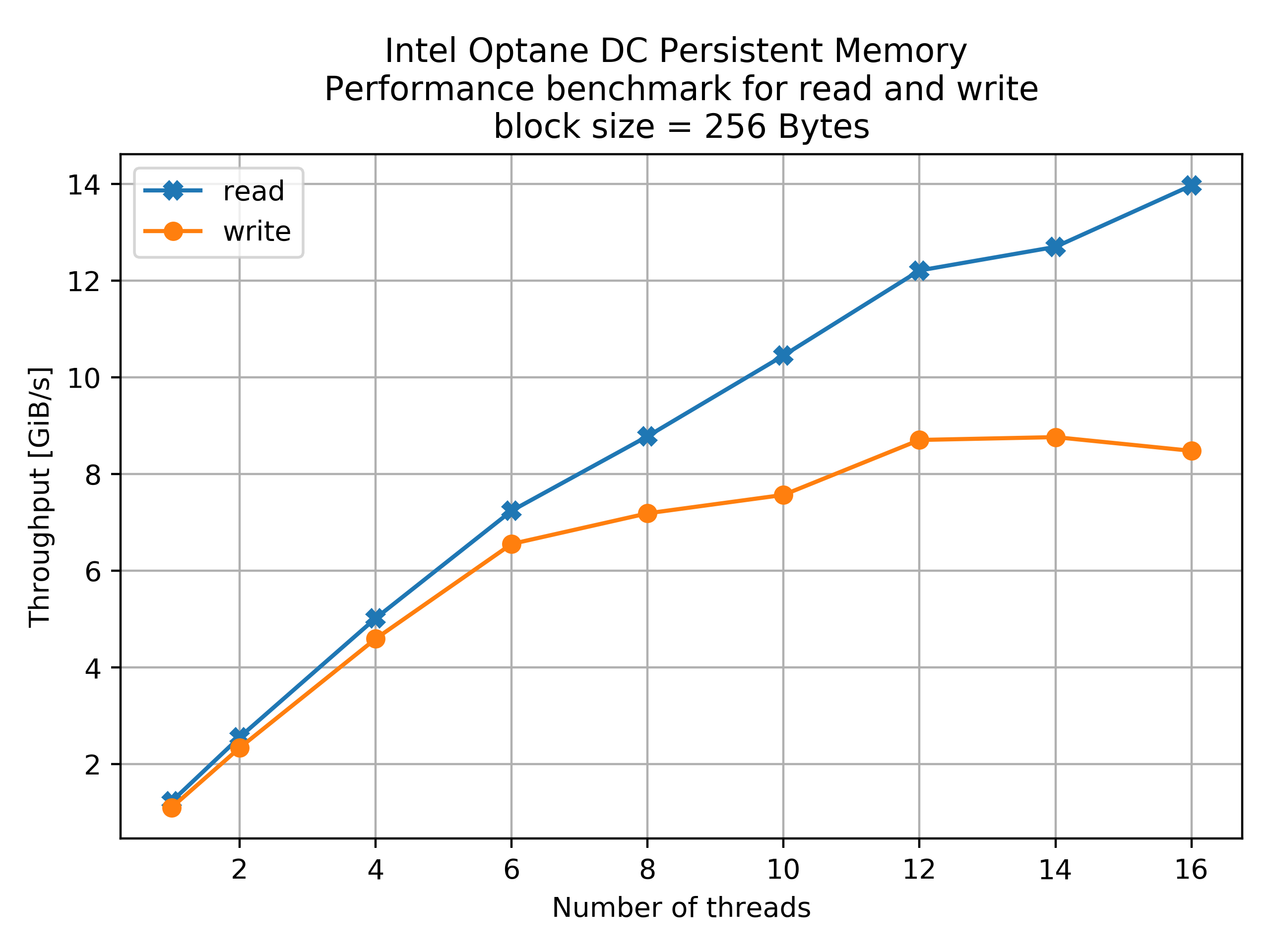}}
\caption{Throughput as a function of number of threads for a block size of 256 bytes in the case of both sequential reading and sequential writing. The maximum write throughput obtained from the application is approximately 8.5~GiB/s whereas for reading a plateau has not been reached even with 16 threads.}
\label{fig:throughput_threads}
\end{figure}

\begin{figure}[t]
\centerline{\includegraphics[width=3.5in]{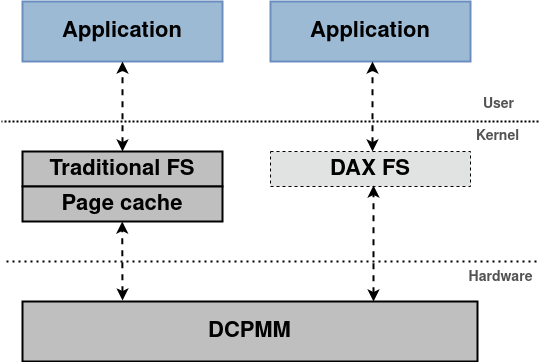}}
\caption{Schematic view of an application running on DCPMMs with a traditional file system and with a DAX-enabled file system.}
\label{fig:pmem_computer_architecture}
\end{figure}

Fig. \ref{fig:throughput_threads} shows the throughput as a function of the number of threads in the case of both the reading and writing access pattern. The block size used for the benchmark is 256 bytes because it represents the lowest access granularity for DCPMM devices. These results were obtained by executing a writing thread on the DCPMMs using an interleaved region and measuring the time taken to write the selected data block size. Note that CPU affinity was set up on the host in order to restrict the executing threads to the physical cores of the same NUMA node of the DCPMMs. This was done to avoid any cross-NUMA access that leads to an increased latency time and, therefore, lower performance. In addition, the writing thread was executed by memory mapping the block of data and then using the MOVNTI \cite{MOVNTI} non-temporal SSE instruction \cite{SSE_instruction} provided by the processor. In this way, the operation has no overhead from the file system because it invalidates the cache line and therefore results in a pure device operation. This feature is possible because it is allowed by DAX-enabled file systems. In this way, it is possible to achieve higher bandwidths. Fig. \ref{fig:pmem_computer_architecture} shows a schematic representation of the difference between running an application on a traditional file system compared to a DAX enabled file system.

From Fig. \ref{fig:throughput_threads} the maximum write throughput obtained from the application is approximately 8.5~GiB/s. In the case of the reading access pattern, the throughput is higher and it has not reached a plateau even when testing with 16 threads. This is because DCPMMs are memory devices that have lower reading latencies compared to the writing operation. Therefore, it is possible to achieve higher bandwidths. As shown in \cite{complete_san_jose}, in the best case configuration, the maximum achieved throughput for a read workload is approximately 40~GiB/s.

\begin{figure}[t]
\centerline{\includegraphics[width=3.5in]{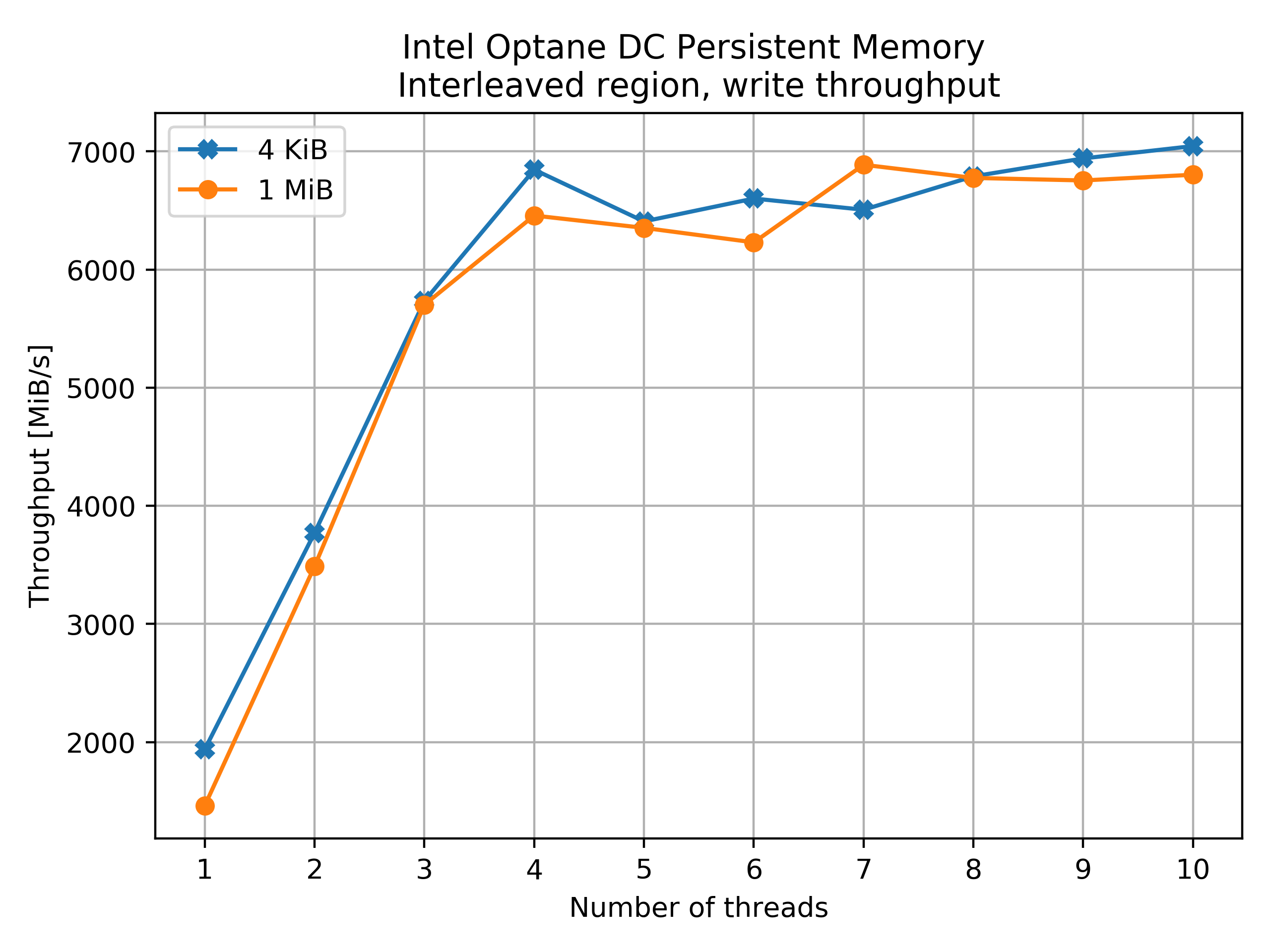}}
\caption{Throughput as a function of the number of threads for both 4~KiB and 1~MiB block sizes. Very limited variability is observed because of the memory-like behaviour of DCPMMs.}
\label{fig:throughput_threads_4k_1M}
\end{figure}

Another interesting feature of DCPMMs is the independence with the access block size. This is illustrated in Fig. \ref{fig:throughput_threads_4k_1M} which shows the writing throughput as a function of the number of threads for two block sizes, respectively 4~KiB and 1~MiB. It can be noted there is less than 5~$\%$ variability between a block size of 4~KiB and 1~MiB. This behaviour is typical of memory devices and it confirms again the memory-like behaviour of DCPMMs.

\section{Application for the DUNE data acquisition system and discussion}

\begin{figure}[t]
\centerline{\includegraphics[width=3.5in]{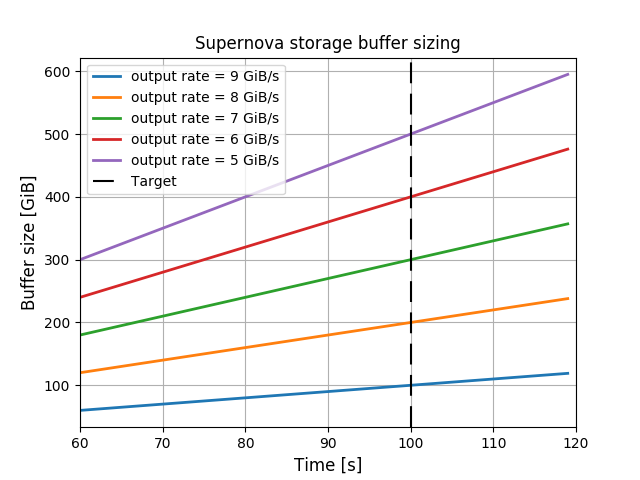}}
\caption{Memory buffer size as a function of the data taking time for different output rates. Decreasing the output rate requires an increased memory size. A line representing the target data taking time of 100~s is also included.}
\label{fig:SNB_storage_size}
\end{figure}

One of the research goals of the DUNE experiment is to detect neutrino signals originating from astrophysical sources such as SuperNova Burst events (SNB). In the data acquisition system of the DUNE detector, data is continuously stored in memory in circular buffers for 10~s before being transferred to permanent storage. Each readout unit generates approximately 10~GB/s, per CPU socket, from 10 separate threads and, upon receiving a trigger signal, data needs to be stored for 100 seconds. This means that the DUNE data acquisition system needs a storage technology capable of sustaining a writing rate of approximately 10~GB/s. Fig. \ref{fig:SNB_storage_size} shows how the memory buffer would need to increase as a function of time depending on different writing bandwidths (output rate) of storage technologies. As an example, if the the writing bandwidth of the selected storage technology is only 5~GiB/s this means that the total memory buffer of the system needs to be increased by 500~GiB in order to keep up with the input rate and for the total time of 100~s. As a consequence, it is necessary to find a suitable storage technology that is capable of sustaining the target rate of 10~GiB/s in order to minimize the extra memory needed to keep the data.

\begin{figure}[t]
\centerline{\includegraphics[width=3.5in]{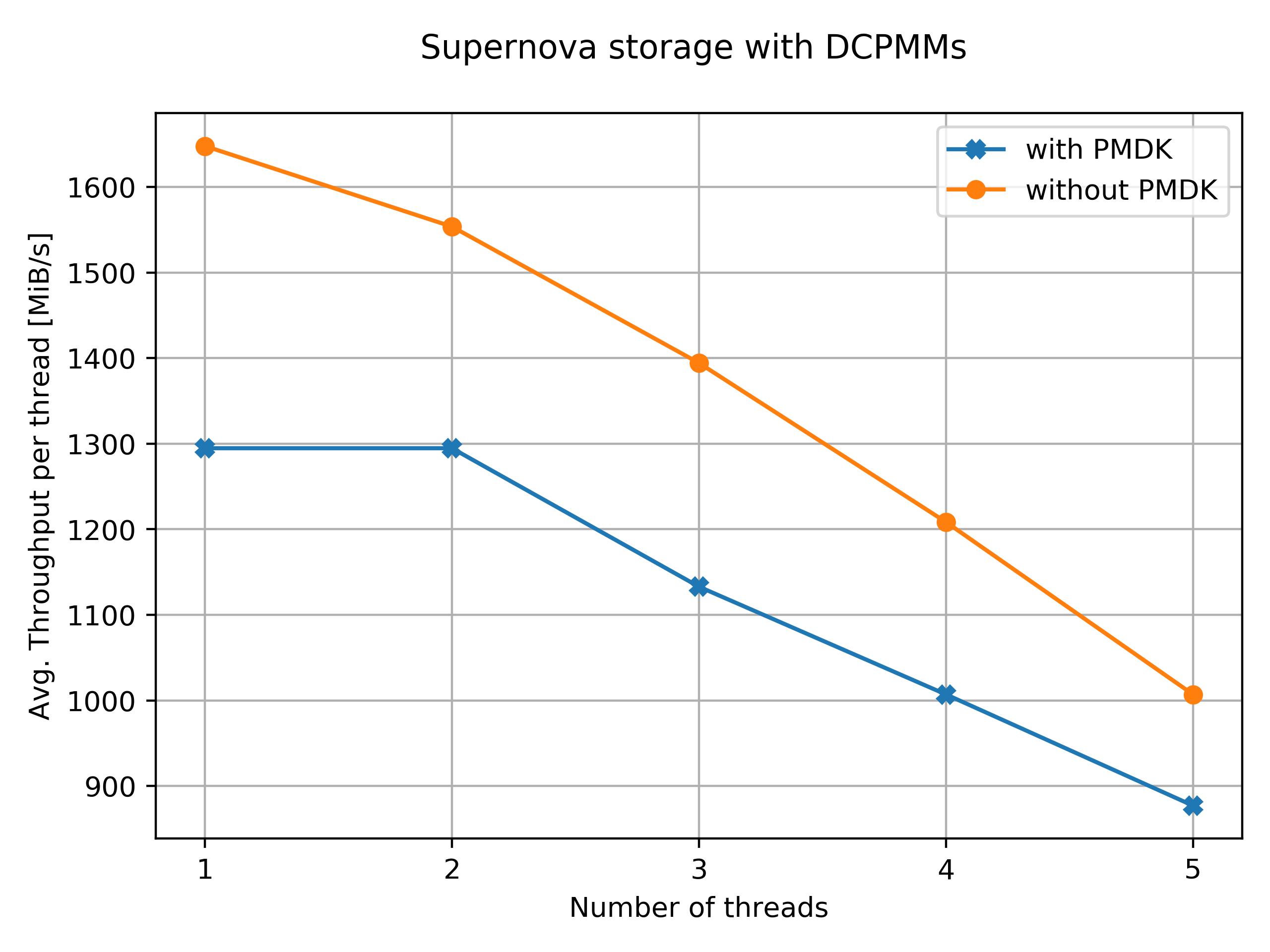}}
\caption{Average throughput per thread as a function of the number of threads for the ProtoDUNE test application with DCPMMs. Results obtained with and without the PMDK software.}
\label{fig:SNB_PMEM_throughput}
\end{figure}

Based on the synthetic benchmarks obtained in section \ref{benchmarks_DCPMMs}, the DCPMMs represent a good candidate for the DUNE supernova storage system. A test application was developed with the DCPMMs integrated in a DUNE prototype setup which is available at CERN. This was deployed with 6 DCPMMs configured in an interleaved configuration. Fig. \ref{fig:SNB_PMEM_throughput} illustrates the performance obtained with the test application. The figure shows the average throughput for an increasing number of threads and for a particular block size of 5568 bytes. This is the access size used in the prototype setup to transfer and store the data from the custom PCIe devices of the readout system. The test application has been written using the Persistent Memory Development Kit (PMDK) \cite{PMDK} which is a collection of libraries and tools that ease the development of applications that use  persistent memory devices. The result obtained in the test application is satisfactory up to four threads because the average throughput per thread can sustain the target bandwidth of 1~GB/s. However, as the the thread number increases the application cannot keep up with the incoming rate.

This lead us to optimize the software stack of the test application in order to fully exploit the performance provided by DCPMMs. It was noticed that the \textit{libpmemblk} library of the PMDK tool was adding extra overheads to the application because of the provided features such as block level atomicity in case of errors or power failures. Similarly to what has been described in section \ref{benchmarks_DCPMMs} a lower level application was developed by creating memory mapped files and then persisting them using the MOVNTI instruction. In this way, there is no extra overhead from the file system and the performance obtained is higher. By optimizing the software application, the throughput obtained in Fig. \ref{fig:SNB_PMEM_throughput} without the use of PMDK increased by approximately 20~$\%$. 

%




\begin{figure}[t]
\centerline{\includegraphics[width=3.5in]{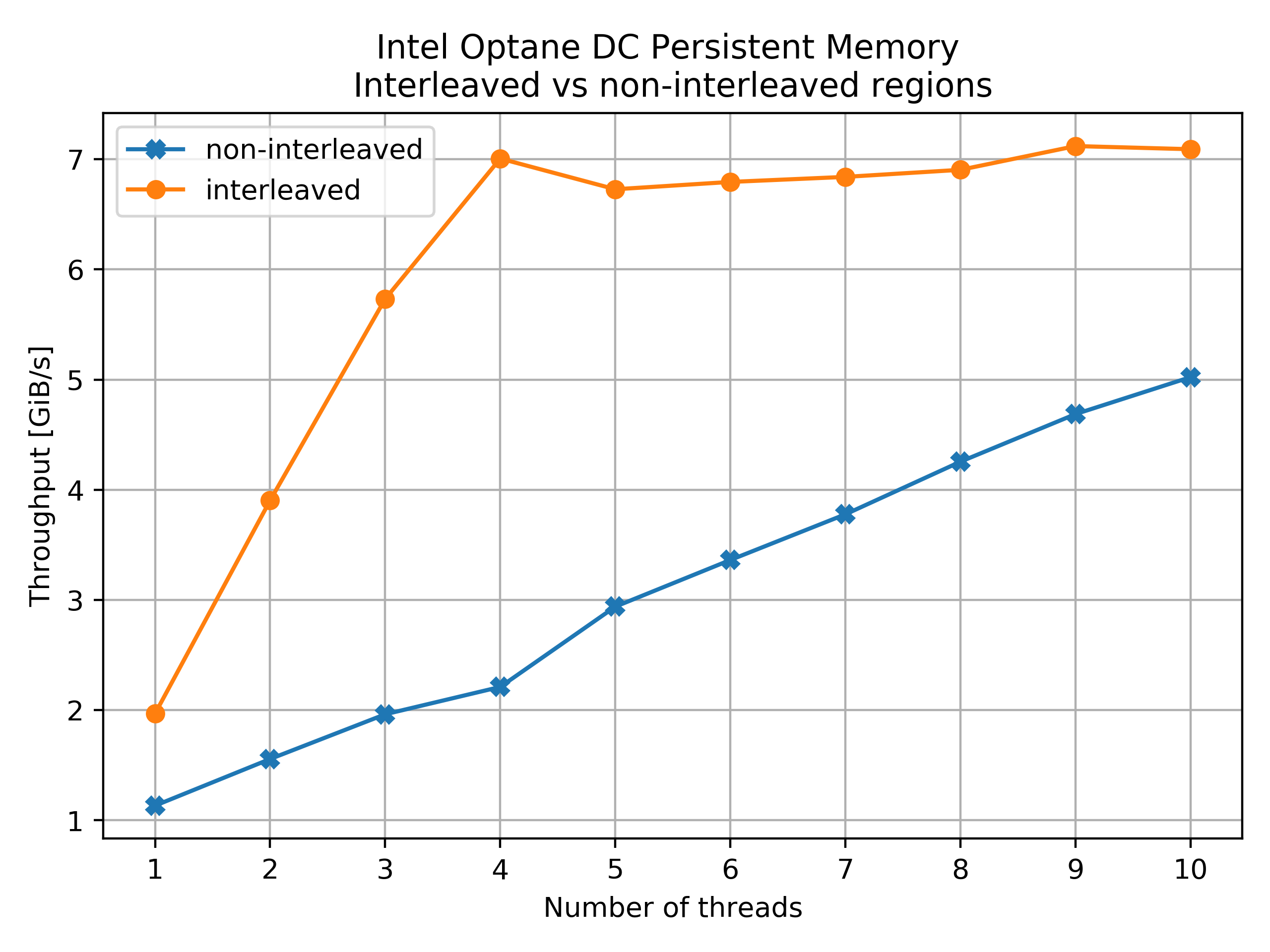}}
\caption{Throughput as a function of number of threads for both interleaved and non-interleaved DCPMM configuration for a DUNE-like application. Maximum throughput is obtained at 7~GiB/s for 4 writing threads.}
\label{fig:superchunk_results}
\end{figure}

A more complete application was developed resembling as much as possible the DUNE workload rather than relying on the current prototype setup. This was obtained by memory copying a static buffer and persisting it with the DCPMMs. All the memory modules available on the system on both sockets have been deployed in the App Direct mode. Fig. \ref{fig:superchunk_results} illustrates the throughput obtained as a function of the number of executing threads for both the interleaved and the non-interleaved DCPMM configuration. The system saturates the available bandwidth with a throughput of 7~GiB/s starting from 4 threads. This means that with the current DCPMMs available today it is possible to sustain, per CPU socket, 80$\%$ of the target throughput required for the DUNE supernova storage buffer. However, the new generation of DCPMMs (Intel\textregistered\, Optane\texttrademark\, Persistent Memory series 200) \cite{PMEM_200} should give on average 25~$\%$ more bandwidth and, therefore, they can fill the required performance that is needed to fully build the DUNE supernova storage buffer.   

The non-interleaved configuration was also tested because it represents a good match for the DUNE supernova workload. In fact, the 10 writing threads required can be considered independent and, therefore, can be configured to write to 10 different block devices. However, as shown in Fig. \ref{fig:superchunk_results} the throughput obtained in this operational mode is much lower. In the case of 5 writing threads the throughput in the non-interleaved configuration is approximately 60$\%$ lower than in the corresponding interleaved DCPMM region. This behaviour confirms that the interleaved configuration is the most suitable for high performance applications and it suggests that DCPMMs in this configuration have internal mechanisms to optimally balance the I/O operations to achieve the best performance.

\section{Related work}
Several research institutes have already tested the Intel\textregistered\, Optane\texttrademark\, DCPMMs in different configurations. Some of the most recent and complete research papers can be found in \cite{complete_san_jose} and in \cite{recent_results_on_PMEM}. 

In our research, the goal is to assess the performance of DCPMMs from an application point of view. Therefore, we decided not to rely only on low level benchmarking tools but instead we developed a high level application, especially designed for a high throughput use-case, that leverage the DCPMM technology. 

Other research groups have also tested DCPMMs from an application perspective. However, rather than focusing on high throughput applications, the typical use case of DCPMMs that is found in literature is to accelerate, for example, database workloads \cite{pmem_DB} or perform faster graph analytics \cite{pmem_graph}. 

\section{Conclusion}

This work has shown the potential provided by the non-volatile memory technology. Intel\textregistered\, Optane\texttrademark\, DC Persistent Memory modules have been tested in detail with both synthetic benchmarks and with a custom-made high level application. The goal was to understand the different operational modes of the technology and investigate the maximum (write) bandwidth that the DCPMM devices can sustain. The ultimate goal was to assess whether DCPMMs could be a viable technology for the implementation of the DUNE supernova storage buffer.

The synthetic benchmarks executed on the system showed that DCPMMs are capable of sustaining high request rates when used as storage devices (App Direct mode). The bandwidth of the system was measured for both the writing and reading access pattern. It was shown that the maximum achieved writing throughput is approximately 8.5~GiB/s starting from 4 threads. 
In addition, the memory-like nature of the DCPMMs was confirmed by measuring the throughput as a function of two different block sizes, respectively 4~KiB and 1~MiB. It was noticed that the system behaves independently of the access size and, thus, confirms the memory nature of the DCPMMs.

A high level application that leverages DCPMMs was also developed and integrated with a prototype setup in order to validate its potential use for the DUNE supernova storage buffer. It was noticed that the overall performance was affected by the overhead added by the PMDK software library. Therefore, software optimizations that take advantage of the MOVNTI instruction set were included and the resulting performance increased by approximately 20~$\%$. A more realistic application resembling the DUNE supernova workload was developed and it was shown that by using all the DCPMMs in the interleaved configuration it is possible to sustain 80~$\%$ of the required throughput.

Future directions of this work will consist in further optimizing the software application to reduce all the overheads and get the maximum throughput from the DCPMMs. In addition, the new generation of Intel\textregistered\, Optane\texttrademark\ Persistent Memory devices could offer a substantial increase in bandwidth which would make possible the deployment of the DUNE supernova storage buffer with DCPMMs.

\section*{Acknowledgment}

We would like to thank Intel\textregistered\, Corporation for providing the hardware necessary to complete this work which was done within the CERN Openlab framework \cite{openlab}. We thank our colleagues from Intel\textregistered\, for the precious feedback and for the fruitful collaboration. 

\printbibliography

\end{document}